# Metastable magnetization behavior in magnetocaloric $R_6Co_{1.67}Si_3$ (R=Tb and Nd) compounds

Arabinda Haldar[1], Niraj K. Singh[2], K. G. Suresh[1,*] and A. K. Nigam[3]

[1]Magnetic Materials Laboratory, Department of Physics, Indian Institute of Technology Bombay, Mumbai- 400076, India

[2]The Ames Laboratory U. S. Department of Energy, Iowa State University, Ames, Iowa 50011-3020, USA

[3]Tata Institute of Fundamental Research, Homi Bhabha Road, Mumbai- 400005, India

*Abstract*

Magnetic field and time induced steps have been observed in the recently discovered ternary silicide $R_6Co_{1.67}Si_3$. Huge relaxation steps are observed across different loops in the low temperature magnetization isotherms. Giant relaxation present in this system indicates the existence of incubation time to get the saturated moment at a certain field. Measurement protocol sensitive magnetization behavior observed in this system may arise from the strong magnetostructural coupling and/or magnetic frustration. Electrical resistivity and magnetoresistance also reflect the magnetic state of the compound. Magnetocaloric effect is found to be large at temperatures close to the magnetic transition temperature.

*Corresponding author (email: suresh@phy.iitb.ac.in)

## 1. Introduction

$R_6Co_{1.67}Si_3$ phase for R=Gd was first realized as an impurity phase in GdCoSi alloys [1]. This series is found to possess very interesting magnetic behavior from the very few studies made till now. Realization of step behavior in the magnetization [2] and the possibility to use as room temperature magnetic refrigerant materials [3] have made this series important from the point of view of both fundamental interest as well as applications. Unusual magnetization steps and time relaxation of magnetization across the metamagnetic transition have been observed in a few intermetallics and oxides [4-9]. The universality of the physical properties among different classes of such materials has been put forward on the basis of the similarities seen in their magnetic properties [10, 11]. Since there are many possible factors for the steps, detailed studies in this direction are essential. In this context, $R_6Co(Ni)_{1.67}Si_3$ series gives a good opportunity to study the staircase-like magnetization growth with field and time.

Non-stoichiometric $R_6Co_{1.67}Si_3$ has been found to possess hexagonal structure (space group $P6_3/m$). This series of compounds has a short *c*- axis, which is reported to produce strain in the system [12]. Magnetization steps at very low temperatures have been observed in $Tb_6Co_{1.67}Si_3$ and $Nd_6Co_{1.67}Si_3$ compounds [2, 13]. We have recently shown that the origin of supercooling effect, phase co-existence and metastability observed in $Ce(Fe_{0.95}Si_{0.05})_2$ compound is due to the strong magneto-structural coupling and the martensitic type behavior of the transition [9]. A part of our attempts to investigate such systems, we have selected a few materials from $R_6Co_{1.67}Si_3$ series and in this paper we report the results obtained in $Nd_6Co_{1.67}Si_3$ and $Tb_6Co_{1.67}Si_3$. We have characterized these compounds using magnetic and electrical resistivity measurements. Time dependent measurements have also been performed to understand the dynamics of the phase transformation. Observed anomalous properties in these two compounds have been compared with those of similar materials.

## 2. Experimental Details

The polycrystalline sample of $Tb_6Co_{1.67}Si_3$ and $Nd_6Co_{1.67}Si_3$ were prepared by arc melting the stoichiometric proportion of the constituent elements of at least 99.9% purity, in a water cooled copper hearth in purified argon atmosphere. The resulting ingots were turned upside down and remelted several times to ensure homogeneity. The as-cast sample was sealed in evacuated quartz tube and annealed for 30 days at 800 °C. The structural analysis of the samples was performed by collecting the room temperature powder x-ray diffraction patterns (XRD) using Cu-K$_α$ radiation. DC magnetization (*M*) measurements were performed in a Squid magnetometer (MPMS-XL, Quantum Design), in vibrating sample magnetometer (VSM) attached to a Physical Property Measurement System (PPMS, Quantum Design) and in a SQUID-VSM. The relaxation measurements were carried out in MPMS SQUID VSM, (Quantum Design). Magnetization measurements have been taken in both zero field cooled (ZFC) and field cooled warming (FCW) modes. Electrical resistivity measurements were carried out in PPMS.

## 3. Results and Discussion

The structural analysis was carried out using x-ray diffraction technique at room temperature (Fig. 1). The refinement of the diffraction patterns was done by the Rietveld analysis using *Fullprof* suite program. The XRD pattern confirms the formation of the desired stoichiometry, with unavoidable secondary phases [14,15]. Refinement shows the existence of Tb$_5$Si$_3$ (12.8 %) and negligible TbCoSi (2.2 %) phases, as also reported earlier [13-15]. The lattice parameters calculated from the refinement are found to be *a* = *b* = 11.6834(4) Å and *c* = 4.1315(2) Å. The refinement results of Tb$_6$Co$_{1.67}$Si$_3$ are shown in Table 1. Refinement of Nd$_6$Co$_{1.67}$Si$_3$ alloy was carried out by taking Nd$_5$Si$_3$ as a secondary phase which turn out to be present in negligible amount (2.2 %). The lattice parameters are calculated to be *a* = *b*= 11.9652(9) Å and *c* = 4.2397(4) Å. The refinement results for Nd$_6$Co$_{1.67}$Si$_3$ are shown in Table 2. In both the compounds, the lattice parameters are in agreement with the reported values [14,15].

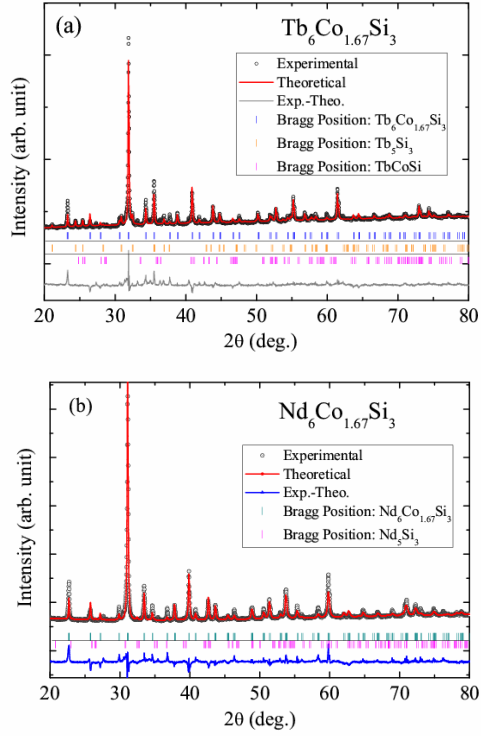

**Fig. 1.** Powder X-ray diffraction patterns, along with the Rietveld refinement of (a) $Tb_6Co_{1.67}Si_3$ and (b) $Nd_6Co_{1.67}Si_3$. The difference plot is shown at the bottom.

**Table 1.** Atomic parameters of $Tb_6Co_{1.67}Si_3$ fully refined by Rietveld refinement technique. The major phase (85.0±0.6%) is $Tb_6Co_{1.67}Si_3$ with space group is $P6_3/m$. The unit cell dimensions are: $a=b=$ 11.6834(4), $c=$ 4.1315(2) Å. Profile residual ($R_P$) and Bragg residual ($R_B$) parameters are 1.93% and 3.62%, respectively.

| Atom | Site | X | y | z | Occupancy |
|---|---|---|---|---|---|
| Tb1 | 6h | 0.7604(7) | -0.0165(8) | 1/4 | 0.500 |
| Tb2 | 6h | 0.1288(7) | 0.5181(7) | 1/4 | 0.500 |
| Co1 | 2b | 0.0000 | 0.0000 | 0.0000 | 0.167 |
| Co2 | 2c | 1/3 | 2/3 | 1/4 | 0.167 |
| Co3 | 4e | 0.0000 | 0.0000 | 0.1692(45) | 0.333 |
| Si | 6h | 0.3314 | 0.1437(24) | 1/4 | 0.500 |

**Table 2.** Atomic parameters of $Nd_6Co_{1.67}Si_3$ fully refined by Rietveld refinement technique. The major phase (97.8±0.4%) is $Nd_6Co_{1.67}Si_3$ with space group is $P6_3/m$. The unit cell dimensions are: $a=b=$ 11.9652(9), $c=$ 4.2397(4) Å. Profile residual ($R_P$) and Bragg residual ($R_B$) parameters are 2.84% and 4.08%, respectively.

| Atom | Site | x | y | z | Occupancy |
|---|---|---|---|---|---|
| Nd1 | 6h | 0.7599(5) | -0.0215(6) | 1/4 | 0.500 |
| Nd2 | 6h | 0.1342(5) | 0.5217(5) | 1/4 | 0.500 |
| Co1 | 2b | 0.0000 | 0.0000 | 0.0000 | 0.167 |
| Co2 | 2c | 1/3 | 2/3 | 1/4 | 0.167 |
| Co3 | 4e | 0.0000 | 0.0000 | 0.1376(33) | 0.333 |
| Si | 6h | 0.3314 | 0.1399(20) | 1/4 | 0.500 |

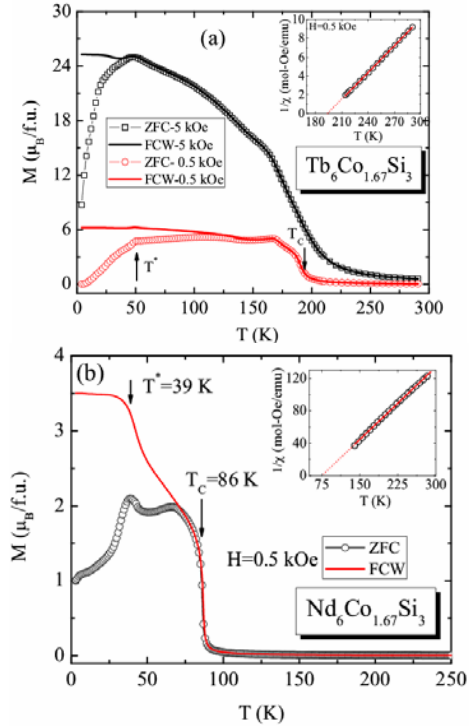

**Fig. 2.** Temperature dependence of magnetization of (a) $Tb_6Co_{1.67}Si_3$ at 500 Oe and 5 kOe and (b) $Nd_6Co_{1.67}Si_3$ at 1 kOe. Open circles and solid line denote ZFC and FCW data respectively. Insets show the inverse susceptibility *vs.* temperature plots and the fit with Curie-Weiss formula.

Temperature dependence of dc magnetization is shown in Fig. 2. The Curie temperature ($T_C$) of $Tb_6Co_{1.67}Si_3$ is found to be 194 K. Another peak is observed at $T^*$=49 K, which is close to the Neel temperature of $Tb_5Si_3$ phase. Below 50 K, the ZFC and FCW data differ considerably. This thermomagnetic irreversibility is significant even in a field as high as 5 kOe and is attributed to the large magnetic anisotropy of the material. Susceptibility data above the Curie temperature can be fitted to Curie-Weiss law, $\chi = C/(T-\theta)$. From the fitting (Inset of Fig. 2a), the paramagnetic Curie temperature is estimated to be 194 K. The effective magnetic moment ($\mu_{eff}$) is found to be 9.1 $\mu_B$/Tb, which is close to the free ion value of 9.7 $\mu_B$ of $Tb^{3+}$ ion. The small difference may be attributed to the secondary phase or to a small moment on Co. These values match well with the earlier reports [13, 14]. $Nd_6Co_{1.67}Si_3$ has $T_C$=86 K and $T^*$=39 K, in agreement with the values reported earlier [14]. It is noteworthy that though the impurity content was much less in this compound, the M-T behavior is almost identical to that of $Tb_6Co_{1.67}Si_3$. The effective moment and the paramagnetic Curie temperature are found to be 3.6 $\mu_B$/Nd and 75 K respectively, as calculated from the inverse susceptibility (inset of Fig. 2b). In this case, the value of $\mu_{eff}$ closely matches with the calculated $g_J[J(J+1)]^{1/2}$ value of free $Nd^{3+}$ ion, implying no moment contribution from Co.

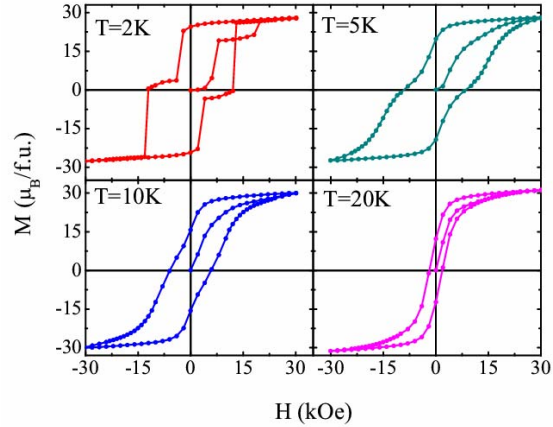

**Fig. 3.** Magnetization isotherms of $Tb_6Co_{1.67}Si_3$ at various temperatures. The sample was zero field cooled from the paramagnetic regime each time.

Fig. 3 shows the isothermal magnetization of $Tb_6Co_{1.67}Si_3$ at different temperatures. Between successive measurements the sample was zero field cooled from temperatures well above the Curie temperature. Five loop magnetization data has been measured at $T$=2, 5, 10 and 20 K, up to a maximum field of 30 kOe. Interestingly, at 2 K, extremely sharp magnetization steps are observed in different quadrants of the hysteresis loop. This feature is quite unique among the intermetallics compounds since the steps are found not only in the 1$^{st}$ or 2$^{nd}$ loops, but in other loops as well. As can be seen, multiple steps are observed in the virgin curve and also the virgin curve lies outside the envelope curve in the high field region. These features are characteristic of supercooling associated with the first order magnetic transition between antiferromagnetic (AFM) and ferromagnetic (FM) phases. A similar of behavior in $M(H)$ isotherm is also observed in doped $CeFe_2$ compound across the field induced antiferromagnetic to ferromagnetic phase transition [6, 8, 9]. However, unlike doped $CeFe_2$, in the present case, there is a large remanence at low temperatures. Therefore, it is quite clear that the thermomagnetic irreversibility seen in Fig. 2 is indeed due to the large anisotropy. With the increase in temperature, the magnetization steps become smoother and the area of the hysteresis loop also decreases. The saturation magnetic moment ($M_S$) has been calculated from the linear extrapolation of $M$ vs. $1/H$ data at high fields. At 2 K, $M_S$ is ~5 $\mu_B$/Tb, which is much smaller than the $g_J J$ value of 9$\mu_B$ for $Tb^{3+}$ ion. This may be due to the crystalline electric field effect and/or a canted magnetic structure at low temperatures. Furthermore, some kind of geometric frustration cannot be ruled out, as in this series of intermetallics $c$-axis is considerably short which induce a strain resulting in Co stoichiometry to be less than 2 [2,12]. In the case of $Nd_6Co_{1.67}Si_3$, at 5 K, the saturation magnetization value is ~1.9 $\mu_B$/Nd, smaller than the $g_J J$ value 3.28 $\mu_B$.

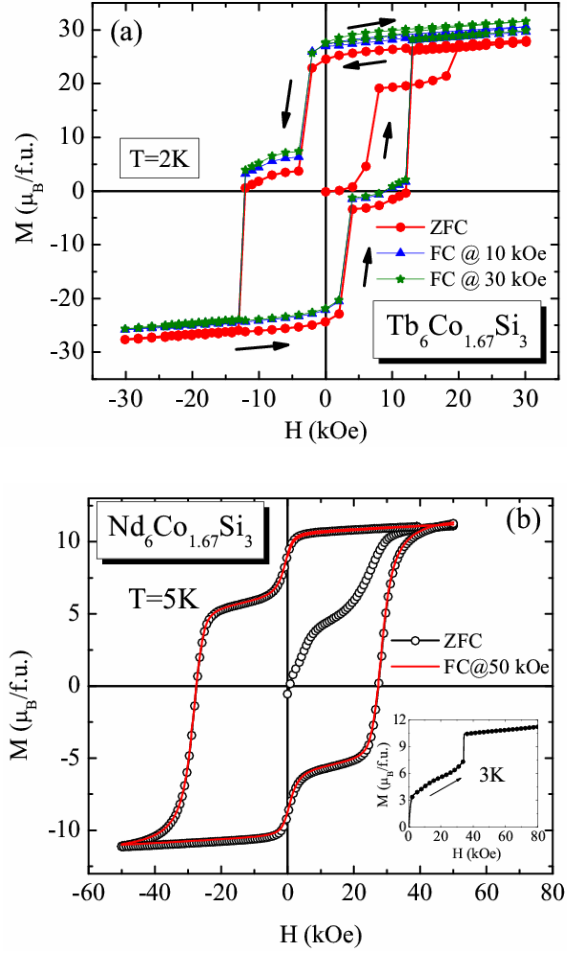

**Fig. 4.** Magnetization isotherm of (a) Tb$_6$Co$_{1.67}$Si$_3$ at 2 K and (b) Nd$_6$Co$_{1.67}$Si$_3$ at 5 K, after zero field cooling and cooling in different fields. The inset in (b) shows the magnetization step at 3 K.

The five loop magnetization data has been collected at low temperatures, after field cooling, as shown in Fig. 4a and b for Tb$_6$Co$_{1.67}$Si$_3$ and in Nd$_6$Co$_{1.67}$Si$_3$ respectively. The data corresponding to the ZFC mode is also shown for comparison. The notable difference between the ZFC and the FC plots in both the compounds is that the virgin curve starts with the saturated moment, in the latter case. But there is almost no difference found in other quadrants in both the compounds. As in the ZFC mode, the sample cooled in the FC mode also shows the magnetization steps. It is to be noted that the step size and the location are the same in both the modes except in the 1$^{st}$ loop. Furthermore, there is negligible difference between the plots obtained after cooling in 20 and 30 kOe, in $Tb_6Co_{1.67}Si_3$. In $Nd_6Co_{1.67}Si_3$ also, though the steps are not sharp at 5 K, the isotherm at 3 K shows a very sharp step (in the virgin path) as seen in the inset of Fig. 4b. The common features seen in both these compounds almost rule out the possibility of any major influence of the secondary phase in the magnetization steps seen in $Tb_6Co_{1.67}Si_3$, rather it shows the general properties of this series, thereby highlighting the similarities with the magnetization behavior of materials with strong magnetostructural coupling. Such systems are generally known to have the coexistence of two competing structural and magnetic phases and the associated metastability of one phase compared to the other. [4-9].

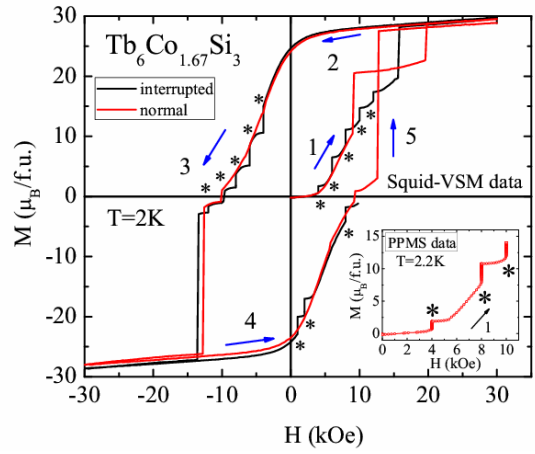

**Fig. 5.** Magnetization isotherm during normal ZFC mode and in the interrupted sweep mode at 2 K. The star positions refer to those fields where the measurement was delayed for 80 minutes. The inset shows the data collected using PPMS VSM.

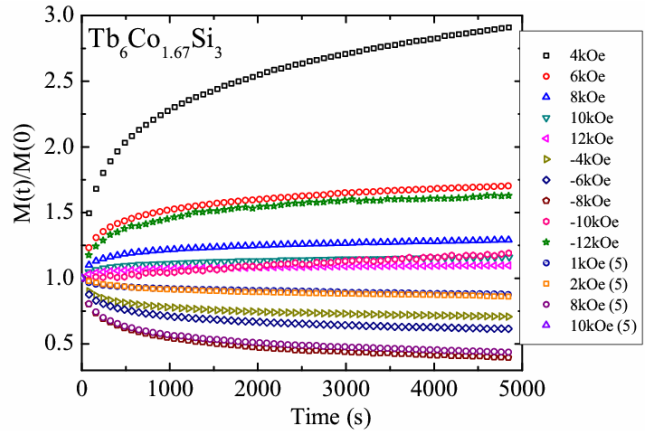

**Fig. 6.** Time evolution of magnetization for $Tb_6Co_{1.67}Si_3$ at 2K when the sample was held for 80 minutes at different fields in $1^{st}$, $3^{rd}$ and $5^{th}$ loops.

The step behavior and the hysteresis have been attributed to the characteristic features of the first order phase transition [8]. As mentioned earlier, usually the first order transitions give rise to coexistence of two different phases, with one of them being metastable. In general, slow magnetization relaxation is a characteristic of metastable systems [16]. Therefore, to find the relaxation time scales, magnetization data was recorded in the interrupted sweep mode as was reported earlier in Ga doped $CeFe_2$ compound and in manganites [8, 17, 18]. In this mode, the field was held constant for 80 minutes at selected fields at different locations of the *M-H* loop. This measurement has revealed quite a few new observations, as evident from Fig. 5 for $Tb_6Co_{1.67}Si_3$. Firstly, it may be noted that one can induce steps by waiting for long durations, both in the increasing and the decreasing field ramps and even in the negative field. Secondly the size of the steps is fairly large which indicates that the compound is highly metastable across the field induced transition region. The maximum change in moment ($\Delta M^{max}$) is found to be 13.2 $\mu_B$/f.u., for a waiting time of 80 minutes. Thirdly, the critical fields where the steps arise are found to change with the measurement protocol. The growth of the reduced moment during the waiting period is shown in Fig. 6. The sluggish relaxation in this material indicates that the material undergoes a field induced magnetic transition through highly metastable states and that the field-induced transition is of first order type. Across the transition region the magnetic states evolves slowly compared to the field sweep rate and while waiting at certain fields the magnetization increases spontaneously. Due to this mismatch of time scale between the relaxation time of the material and the experimental sweep rate, one sees spontaneous magnetization jumps on waiting. It was found that the relaxation data cannot be fitted to that of typical spin glass relaxation rate. This rules out spin glass type behavior at low temperature, which is in agreement with the frequency dependence of the ac susceptibility data [13].

At this point, it is worth mentioning the differences in the data collected using different magnetometers. We found a clear difference in the data collected using Squid-VSM and the conventional squid magnetometer. On the other hand, the Squid-VSM data matches well with the PPMS VSM data [13]. The data collected by us using the PPMS VSM is shown in the inset of Fig. 5a. It should be kept in mind that in conventional squid, the measurement is done in the scan mode in which the field is stabilized before taking the data at that point, while VSM generally uses the sweep mode. Therefore, the differences in the data observed in different magnetometers indicate the protocol dependence of magnetization in this system. Such dependence is a characteristic property of systems with slow kinetics, like the martensitic systems [11, 19].

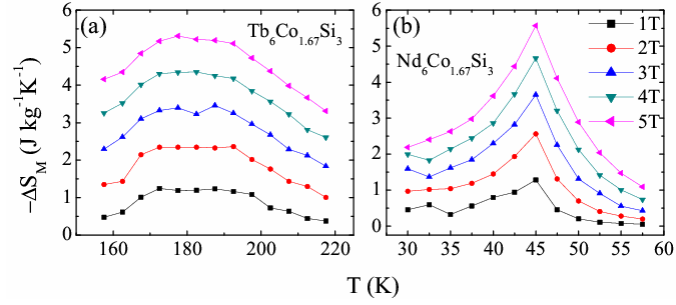

**Fig. 7**. Magnetic entropy change $(-\Delta S_M)$ vs. temperature plot for (a) $Tb_6Co_{1.67}Si_3$ and (b) $Nd_6Co_{1.67}Si_3$ compound.

Magnetic entropy change (Fig. 7) which is a measure of magnetocaloric effect (MCE) has been calculated using Maxwell's relations [20],

$$\Delta S_M(T, \Delta H) = \int_{H_1}^{H_2} \left( \frac{\delta M(T,H)}{\delta T} \right)_H dH \quad (1)$$

With the *M(H)* isotherms taken at constant temperature intervals, the above relation can be approximated to the following expression:

$$\Delta S_M \approx \frac{1}{\Delta T} \left[ \int_{H_1}^{H_2} M(T+\Delta T, H) dH - \int_{H_1}^{H_2} M(T, H) dH \right] \quad (2)$$

Magnetization isotherms have been taken at various temperatures around transition region at 5 K interval. For $Tb_6Co_{1.67}Si_3$ compound a broad peak is observed around 190 K (Fig. 7(a)) in the temperature dependence of the magnetic entropy change plot. On the other hand, a sharp peak is found in $Nd_6Co_{1.67}Si_3$ compound (Fig. 7(b)). The difference in the shape of *ΔS$_M$* vs. *T* plots of two compounds reflects the $(T_C - T^*)$ value in these two cases. Maximum value of the entropy change $(-\Delta S_M^{max})$ is ~5.5 J kg$^{-1}$ K$^{-1}$ for both these compounds for a field change of 50 kOe. The maximum MCE value in the case of $Gd_6Co_{1.67}Si_3$ is 5.5 J kg$^{-1}$ K$^{-1}$ for a field change of 48 kOe. [3].

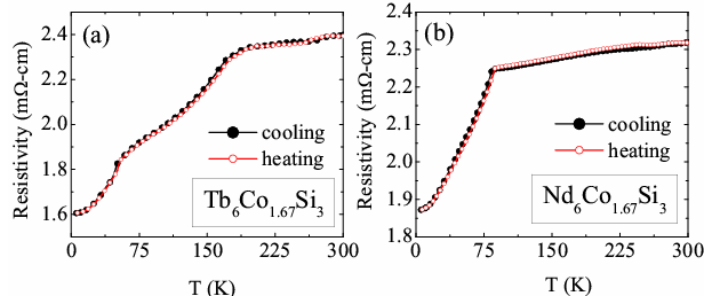

**Fig. 8.** Temperature dependence of resistivity for (a) $Tb_6Co_{1.67}Si_3$ and (b) $Nd_6Co_{1.67}Si_3$ in zero field (data taken in cooling and heating modes) and at 5 kOe in FCW mode.

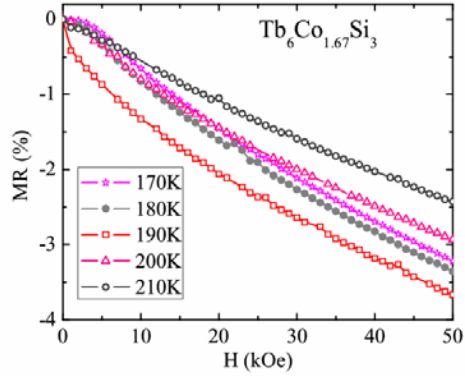

**Fig. 9.** Variation of magnetoresistance with field at different temperatures (around the transition region) for $Tb_6Co_{1.67}Si_3$.

Temperature dependence of electrical resistivity has been studied in zero field and in a field of 5 kOe (Fig. 8). There is no significant difference between these two modes, though there was considerable difference in the magnetization data. Martensitic systems such as magneto-resistive oxides and doped $CeFe_2$ are known to show large difference between the field cooled and zero field cooled resistivity data [16, 21]. As can be seen from Fig. 8, the resistivity data of $Tb_6Co_{1.67}Si_3$ shows an anomaly corresponding to the magnetic transition at about 190 K, as seen in the *M-T* data. It shows an additional anomaly at about 50 K (Fig. 8(a)), which is also present in the *M-T* data. But no extra peak other than the one at 86 K is observed in $Nd_6Co_{1.67}Si_3$ (Fig. 8(b)). Magnetoresistance (MR) has been calculated using the relation $[(\rho(H) - \rho(0))/\rho(0)]$ for $Tb_6Co_{1.67}Si_3$ (Fig. 9). The maximum MR is calculated to be -3.7% at around 190 K. It is found that in general, the MR shows a quadratic dependence on the field, except at very low fields, which implies that spin fluctuations play a major role in contributing to MR at these temperatures [22].

The results presented above show that there are some common features between the martensitic systems reported earlier. Another possibility is the role of quenched disorder, which is known to result in similar anomalies [23]. At this point, it is not possible to identify the exact reason in the present case. However, we feel that the smaller $c$-axis and vacancies in the Co sublattice, as reported by Gaudin *et al.* [12], has some role in the observed properties. The trigonal arrangement of rare earth ions in the unit cell may cause frustration in this system. Another point that comes out of this study is that though there is a larger impurity in $Tb_6Co_{1.67}Si_3$ compared to that in $Nd_6Co_{1.67}Si_3$, the magnetization behavior is more or less identical in both the compounds.

## 4. Conclusions

We have carried out magnetic and resistivity measurements on $Tb_6Co_{1.67}Si_3$ and $Nd_6Co_{1.67}Si_3$ compounds. In addition to the magnetization steps, time evolution of magnetization is particularly emphasized as it reveals fairly large spontaneous magnetization growth. There are many similarities such as sharp magnetization jumps, sluggish relaxation, etc. between the present series and some martensitic systems such as doped $CeFe_2$ and phase separated oxides. Across the field induced transition region from low magnetic state to high magnetic state, the compound is found to be metastable, which results in slow magnetization relaxation. This indicates that the rate of nucleation and growth of the field induced high magnetic state is slow compared to the experimental time scales, resulting in the metastable behavior in the physical properties. The electrical resistivity data is found to be in agreement with the magnetization data, in general. To find out the exact reason for the anomalies, use of local probes such as neutron diffraction, micro Hall imaging etc., is essential.

*Acknowledgements*

KGS and AKN thank BRNS for the financial support for carrying out this work. The authors also thank Mr. Devendra D. Buddhikot for his help in the resistivity measurements.